\documentclass[10pt]{article} % onecolumn (standard format)

\usepackage[utf8]{inputenc}
\usepackage{amsmath}
\usepackage{amsfonts}
\usepackage{amssymb}
\usepackage{mathtools}
\usepackage{listings}
\usepackage{mathrsfs}
\usepackage{times}
\usepackage{float}
\usepackage{color}
\usepackage{setspace}
\usepackage{color,soul}

\usepackage{amsmath,amsfonts,amsthm,eucal}
\usepackage{enumerate}
\usepackage[letterpaper,margin=3cm]{geometry}
\usepackage{float}
\usepackage[title]{appendix}
\usepackage{color}
\usepackage{tabularx}
\usepackage{siunitx}
\usepackage[tableposition=below]{caption}
\usepackage[
pdfstartview=XYZ,
bookmarks=true,
colorlinks=true,
linkcolor=blue,
urlcolor=blue,
citecolor=blue,
pdftex,
bookmarks=true,
linktocpage=true,   % makes the page number as hyperlink in table of content
hyperindex=true
]{hyperref}

\usepackage{lipsum}
\usepackage[FIGBOTCAP,TABTOPCAP,bf,tight]{subfigure}

\usepackage{natbib}
\usepackage[pdftex]{graphicx}
\usepackage{epstopdf}
\usepackage{booktabs} 
\usepackage{tikz,mathpazo}
\usetikzlibrary{shapes.geometric, arrows}
\usepackage{caption}

\newcommand{\tensor}[1]{\ensuremath{\boldsymbol{#1}}}

\DeclareMathOperator{\grad}{\nabla}
\DeclareMathOperator{\symgrad}{\nabla^{\text{s}}}
\DeclareMathOperator{\diver}{\nabla\cdot}

\theoremstyle{remark}

\renewcommand{\vec}[1]{\ensuremath{\boldsymbol{#1}}}

\usepackage{algorithm}
\usepackage{algorithmicx}
\usepackage[noend]{algpseudocode}

\theoremstyle{definition}

\usepackage{longtable}
\usepackage{adjustbox}
\usepackage{framed}

\title{An immersed
phase field fracture model for fluid-infiltrating porous media
with evolving Beavers-Joseph-Saffman condition} 

\begin{document}

\author{Hyoung Suk Suh\thanks{Department of Civil Engineering and Engineering Mechanics, 
 Columbia University, 
 New York, NY 10027.     \textit{h.suh@columbia.edu}  }       \and
        WaiChing Sun\thanks{Department of Civil Engineering and Engineering Mechanics, 
 Columbia University, 
 New York, NY 10027.
  \textit{wsun@columbia.edu}  (corresponding author)      %     \\
}
}

\maketitle

\begin{abstract}
This study presents a phase field model for brittle fracture in fluid-infiltrating vuggy porous media. 
While the state-of-the-art in hydraulic phase field fracture considers Darcian fracture flow with enhanced permeability along the crack, in this study, the phase field not only acts as a damage variable that provides diffuse representation of cracks or cavities, but also acts as an indicator function that separates the domain into two regions where fluid flows are governed by Stokes and Darcy equations, respectively. 
Since the phase field and its gradient can be respectively regarded as smooth approximations of the Heaviside function and Dirac delta function, our new approach is capable of imposing interfacial transmissibility conditions without explicit interface parametrizations. 
In addition, the interaction between solid and fluid constituents is modeled by adopting the concept of mixture theory, where the fluid velocities in Stokes and Darcy regions are considered as relative measures compared to the solid motion. 
This model is particularly attractive for coupled flow analysis in geological materials with complex microstructures undergoing brittle fracture often encountered in energy geotechnics problems, since it completely eliminates the needs to generate specific enrichment function, integration scheme, or meshing algorithm tailored for complex geological features. 
\end{abstract}

\section{Introduction}
\label{intro}
Defects, such as cracks, joints, vuggy pores and cavities and impurities are important for the hydro-mechanical coupling encountered in porous media in energy geotechnics problems like enhanced oil recovery or the development of enhanced geothermal energy reservoirs. 
When the defects are partially or fully filled with pore fluid, the size and the geometric features of the defects may both impose significant changes on both the effective stiffness and permeability of the materials as well as the Biot's coefficient \citep{wang2017anisotropy} and the undrained and drained shear strength \citep{wang2019meta, sun2013unified}. 
For instance, carbonate rocks and limestone often contains pores of profoundly different sizes \citep{ coussy2004poromechanics, sun2011connecting, na2017computational, choo2018cracking, sun2018prediction}.

One possible modeling choice is to not explicitly model each defect and imperfects but instead incorporate the influences of these defects in the constitutive laws of an imaginary effective medium at a scale where a representative elementary volume exists. 
In this case, defects may be treated as a different pore system that may interact with the prime pore space through fluid mass exchanges, as shown in the multi-porosity and multi-permeability models in the literature \citep{liu2017shale, choo2016hydromechanical, wang2019updated, ma2020computational}. 
The upshot of this model is mainly the relatively simple numerical treatment. 
Due to the absence of defects in the homogenized effective medium, there is no need for complex meshing techniques. 
The absence of embedded strong discontinuities or sharp gradient of phase field or level set also make it more feasible to employ standard finite element or finite element/finite volume solver, provided that the numerical stability is ensured \citep{zienkiewicz1999computational, sun2013stabilized, sun2015stabilized}. 
However, the drawback of these approaches is that the homogenized effective medium may not be sufficiently representing the microstructure. 
Furthermore, the identification of material parameters can become more complicated as 1) the effective permeability of the multiple interacting pore systems are generally not anisotropic and not co-axial with each other and 2)  the constitutive law for the fluid mass exchanges are inherently dependent on the geometric attributes such as surface areas and the mechanisms may change over time as cracks grow, coalescence and heal and interact with pores exciting in the porous media.

Explicitly capturing the geometrical evolution of these defects may resolve these issues. 
However, this approach has been an ongoing challenges for the poromechanics community in the past decades \citep{detournay1993fundamentals, wang2018multiscale, choo2018cracking}. 
From the modeling prospective, it is essential to employ a proper representation for the defects either explicitly (e.g. insertion of cohesive zone element, enriching basis function or assumed strain methods) or implicitly (e.g. the phase field, level set or eigenfracture approaches). 
While both approaches have achieved a level of success in the past decades for solid mechanics problems, the modeling effort of the hydraulic responses of defects are often limited to cubic law models that relates hydraulic aperture to hydraulic conductivity \citep{pyrak1988fluid,sisavath2003simple}. 
While the cubic law can be applied to one-dimensional case where local aperture does not fluctuate, the actual flow pattern inside the crack is not necessarily one dimensional. 
In fact, the macroscopic flow pattern inside the crack is also dominated by the crack geometry and hence the cubic law or simply enhancing the permeability is not universally feasible and certainly not appropriate in highly heterogeneous domain.

In this work, our contribution is to introduce the coupling of the Stokes-Darcy flow that enable a unified treatment for flow in the defects. 
By leveraging the phase field not only as an indicator function for the location of cracks but also other defects such as cavities and large or geometrically complicated voids that does not fit for computational homogenization, we introduce a phase field framework that may efficiently couple the Stokes flow in the defect regions that interact with the pore fluid infiltrating in the intact porous matrix while the Stokes and Darcy regions are evolving due to the crack growth. 
By explicitly modeling the fluid-filled defects, we bypass the need to introducing additional phenomenological models for the mass exchanges and enable modelers to capture how different geometries of these defects, meso-scale voids and cracks affect the macroscopic outcome. 
Numerical examples are provided to showcase the potential applications of this proposed models.

\section{Methods}
\label{methods}

\subsection{Diffuse representation of Stokes-Darcy system}
\label{diffuse_representation}
\begin{figure}[h]
\centering
\includegraphics[height=0.375\textwidth]{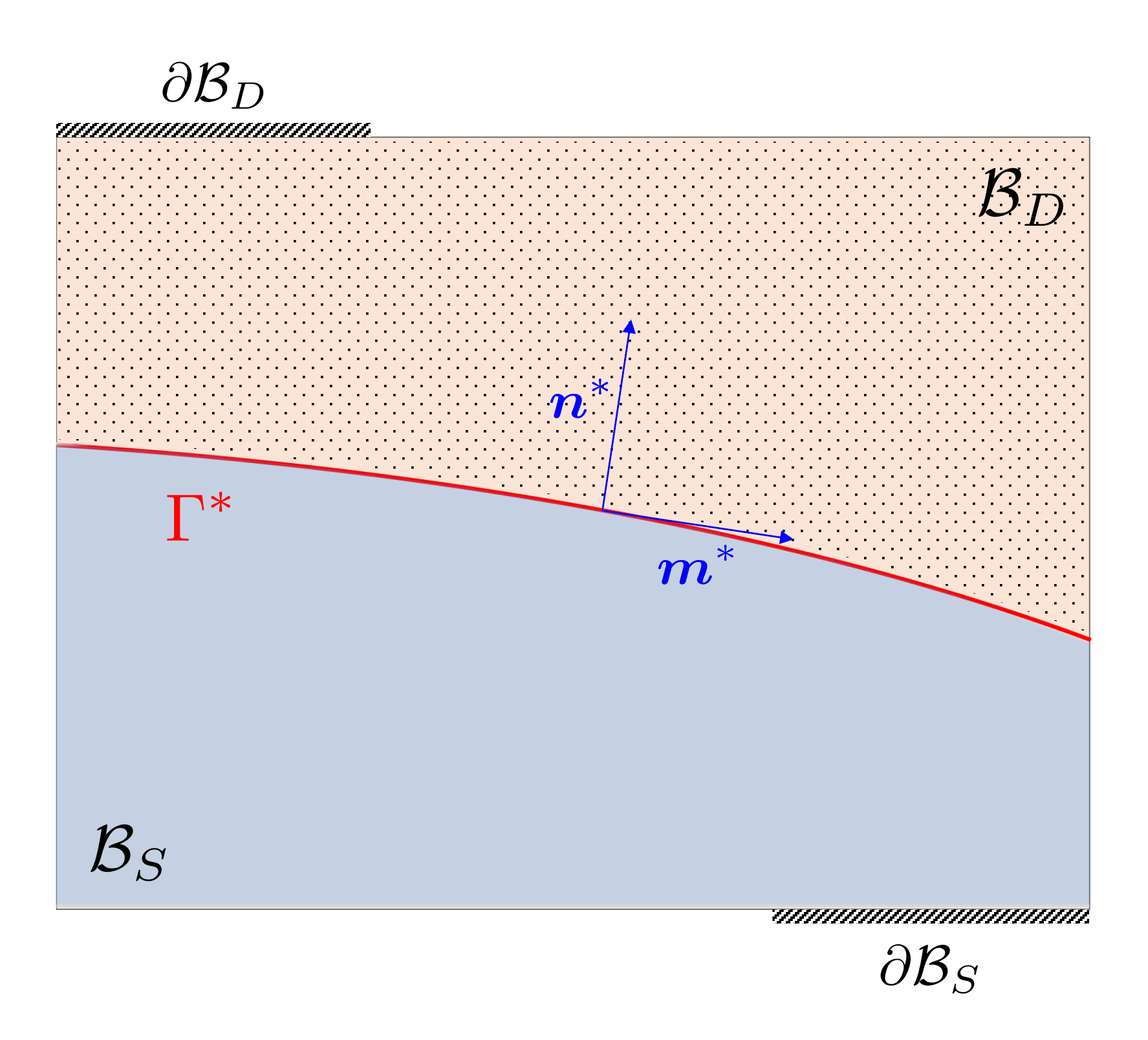}
\caption{Schematic representation of computational domain of interest $\mathcal{B}$. $\mathcal{B}_D$ is the porous matrix while $\mathcal{B}_S$ represents a subdomain for vuggy regions.}
\label{fig:domain_decomposition}
\end{figure}

In this study, we model coupled free and porous medium flow by considering two subdomains (i.e., cracks or vugs $\mathcal{B}_S$ and porous matrix $\mathcal{B}_D$, where fluid flow is governed by Stokes and Darcy's equations, respectively) separated by a sharp interface $\Gamma^*$ (Fig.~\ref{fig:domain_decomposition}). 

\begin{figure}[h]
\centering
\includegraphics[height=0.4\textwidth]{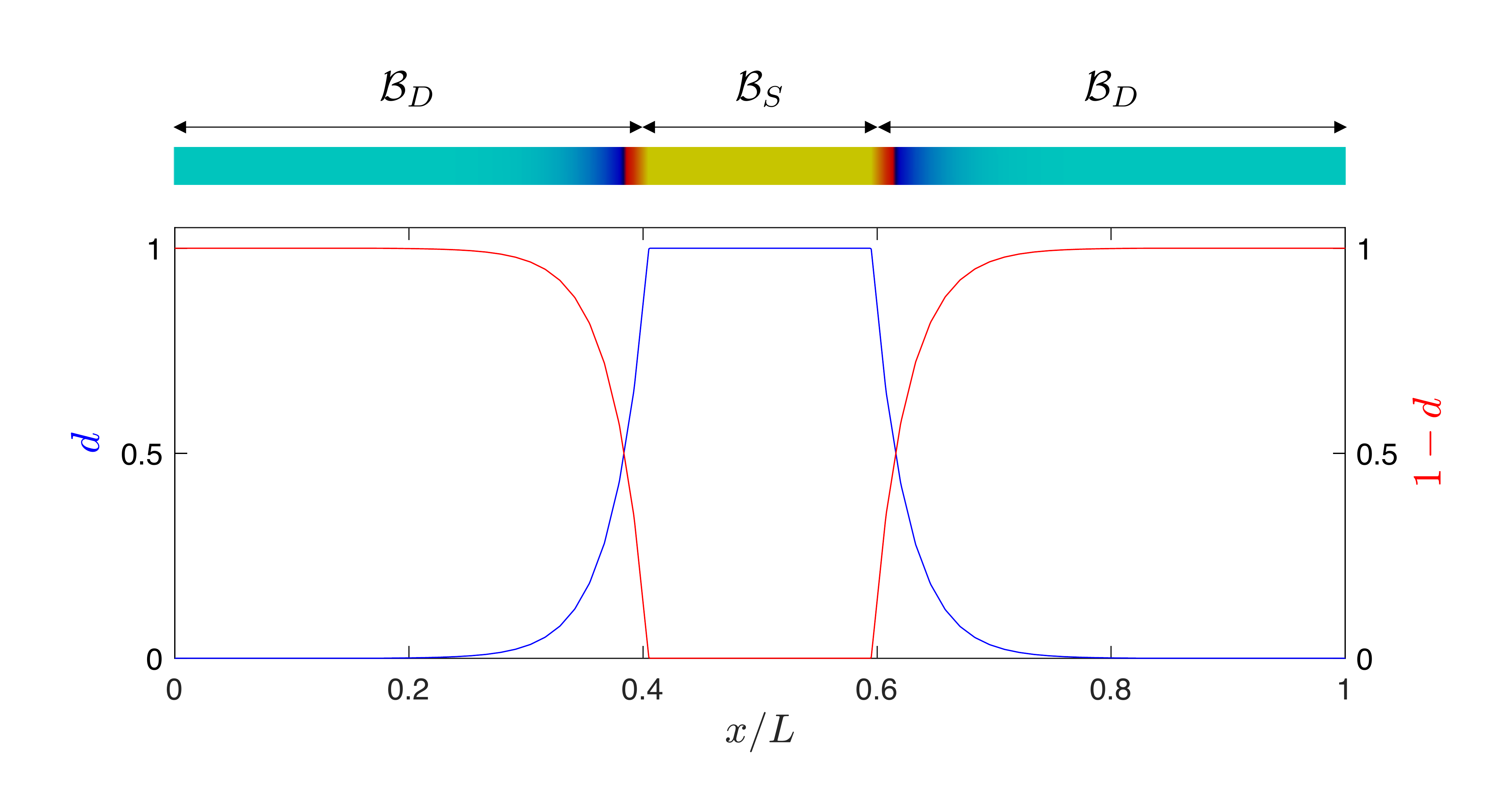}
\caption{Diffuse representation of the interface where exemplary 1D domain consists of Stokes region $x/L \in [0.4, 0.6]$, sandwiched between Darcy regions.}
\label{fig:smooth_representation}
\end{figure}

Similar to \citet{stoter2017diffuse}, we attempt to employ a diffuse approximation for the sharp interface between two regions via implicit function. 
By adopting a phase field approach which is widely used in modeling fracture \citep{miehe2010phase, borden2012phase, suh2019open, SUH2020113181}, we approximate the interfacial area $A_{{\Gamma}^*}$ as $A_{{\Gamma}^*_d}$, which can be expressed in terms of volume integration of surface density ${\Gamma}^*_d (d, \grad{d})$ over $\mathcal{B}$:
\begin{equation}
\label{eq:surface_density}
A_{{\Gamma}^*} \approx A_{{\Gamma}^*_d} = \int_{\mathcal{B}} \Gamma^*_d \left( d, \grad{d} \right) \: dV,
\end{equation}
where $d$ is the phase field which varies from 0 in Darcy region ($\mathcal{B}_D$) to 1 in Stokes region ($\mathcal{B}_S$). 
This study considers the density functional originally used to introduce elliptic regularization of the Mumford-Shah functional \citep{mumford1989optimal} which possesses quadratic local dissipation function, i.e., 
\begin{equation}
\label{eq:crack_density}
\Gamma^*_d \left( d, \grad{d} \right) = \frac{1}{2} \left[ \frac{d^2}{2l^*} + \frac{l^*}{2} \left( \grad{d} \cdot \grad{d} \right) \right].
\end{equation}
Here, $l^*$ is the regularization length that controls the size of diffuse interface. 
By solving a differential equation which can be recovered by seeking the stationary point where the functional derivative of the density functional in Eq.~\eqref{eq:crack_density} vanishes, we can obtain diffuse representation of the Stokes-Darcy system (e.g., Fig.~\ref{fig:smooth_representation}).

Since $\Gamma$-convergence requires that the sharp interface is recovered by reducing the regularization length $l^*$ to zero \citep{bourdin2008variational, amor2009regularized}, 
the phase field and its gradient can be regarded smooth approximations of the Heaviside function $H$ and Dirac delta function $\delta_{\Gamma^*}$, respectively \citep{stoter2017diffuse}. 
Therefore, the volume integral of an arbitrary function $\tilde{G}$ over Stokes and Darcy regions can be approximated as follows:
\begin{equation}
\label{eq:vol_int}
\int_{\mathcal{B}_S} \tilde{G} \: dV = \int_{\mathcal{B}} \tilde{G} H \: dV \approx \int_{\mathcal{B}} \tilde{G} d \: dV
\: \: ; \: \:
\int_{\mathcal{B}_D} \tilde{G} \: dV = \int_{\mathcal{B}} \tilde{G} (1-H) \: dV \approx \int_{\mathcal{B}} \tilde{G} (1-d) \: dV.
\end{equation}
Likewise, the surface integral of the function $\tilde{G}$ over the interface is approximated as,
\begin{equation}
\label{eq:surf_int}
\int_{\Gamma^*} \tilde{G} \: d\Gamma 
=  \int_{\mathcal{B}} \tilde{G} \delta_{\Gamma^*} \: dV
\approx \int_{\mathcal{B}} \tilde{G} \| \grad{d} \| \: dV,
\end{equation}
where the magnitude of the gradient of the phase field approximates a Dirac delta distribution at the sharp interface $\Gamma^*$. 
We can also obtain the normal ($\vec{n}^*$) and tangent ($\vec{m}^*$) vectors (Fig.~\ref{fig:domain_decomposition}) from the phase field representation of the interface. 
The normal vector can be computed as:
\begin{equation}
\label{eq:n_vec}
\vec{n}^* = - \frac{\grad{d}}{\| \grad{d} \|}.
\end{equation}
In the general three-dimensional case, the tangent vectors can then be obtained as,
\begin{equation}
\label{eq:t_vec}
\vec{m}^*_1 =
\begin{dcases}
(n^*_y, -n^*_x, 0) & \text{if } \| n^*_x \| > \| n^*_y \| \text{ and } \| n^*_x \| > \| n^*_z \|, \\
(0, n^*_z, -n^*_y) & \text{otherwise}
\end{dcases}
\: \: ; \: \:
\vec{m}^*_2 = \vec{n}^* \times \vec{m}^*_1.
\end{equation}

\subsection{Governing equations for hydraulic phase field fracture}
\label{governing_eqs}
In this study, the interaction between solid and fluid constituents is modeled by adopting the mixture theory, where the fluid velocities $\bar{\vec{w}}_{fS}$ and $\bar{\vec{w}}_{fD}$ in both regions become relative measures compared to the solid motion. 
For simplicity, we assume that the vuggy porous material is composed of incompressible solid and fluid constituents under quasi-static conditions. 
We also assume that effective stress principle is only valid in $\mathcal{B}_D$ while solid constituent inside $\mathcal{B}_S$ is fully liquefied. 
In this case, we may write local forms of the governing equations as: 
\begin{equation}
\label{eq:stokes}
\begin{dcases}
\diver{ \left( -p_{fS} \tensor{I} + 2 \mu_f \symgrad{\dot{\vec{u}}} \right)} = \vec{0} \:\: &\text{in } \mathcal{B}_S, \\
- \diver{ \left( -p_{fS} \tensor{I} + 2 \mu_f \symgrad{\bar{\vec{w}}_{fS}} \right)} = \vec{f}_{fS} \: \: &\text{in } \mathcal{B}_S, \\
\diver{\dot{\vec{u}}} + \diver{\bar{\vec{w}}_{fS}} = 0 \: \: &\text{in } \mathcal{B}_S, \\
\end{dcases}
\end{equation}
\begin{equation}
\label{eq:darcy}
\begin{dcases}
\diver{ \left( \tensor{\sigma}' - p_{fD} \tensor{I} \right) } = \vec{f} \:\: &\text{in } \mathcal{B}_D, \\
\diver{\dot{\vec{u}}} + \diver{\bar{\vec{w}}_{fD}} = 0 \: \: &\text{in } \mathcal{B}_D,
\end{dcases}
\end{equation}
\begin{equation}
\label{eq:phase_field}
g'(d) \mathcal{H} + \frac{\mathcal{G}_c}{l^*} \left( \frac{d}{2} - \frac{{l^*}^2}{2} \nabla^2 d \right) = 0 \: \: \: \text{in } \mathcal{B},
\end{equation}
%\begin{alignat}{2}
%\label{eq:eq7}
%&\diver{ \left( -p_{fS} \tensor{I} + 2 \mu_f \symgrad{\dot{\vec{u}}} \right)} = \vec{0} \:\: &&\text{in } \mathcal{B}_S, \\
%\label{eq:eq8}
%&\diver{ \left( \tensor{\sigma}' - p_{fD} \tensor{I} \right) } = \vec{f} \:\: &&\text{in } \mathcal{B}_D, \\
%\label{eq:eq9}
%&- \diver{ \left( -p_{fS} \tensor{I} + 2 \mu_f \symgrad{{\vec{w}}_{fS}} \right)} = \vec{f}_{fS} \: \: &&\text{in } \mathcal{B}_S, \\
%\label{eq:eq10}
%&\diver{\dot{\vec{u}}} + \diver{\vec{w}_{fS}} = 0 \: \: &&\text{in } \mathcal{B}_S, \\
%\label{eq:eq11}
%&\diver{\dot{\vec{u}}} + \diver{\vec{w}_{fD}} = 0 \: \: &&\text{in } \mathcal{B}_D, \\
%\label{eq:eq12}
%&g'(d) \mathcal{H} + \frac{\mathcal{G}_c}{l^*} \left( \frac{d}{2} - \frac{{l^*}^2}{2} \nabla^2 d \right) = 0 \: \: &&\text{in } \mathcal{B},
%\end{alignat}
where $\vec{u}$ is the solid displacement; $p_{fS}$ and $p_{fD}$ are fluid pressures in Stokes and Darcy regions, respectively; $\mu_f$ indicates fluid viscosity; $\tensor{\sigma}'$ is the effective stress; and $\mathcal{G}_c$ is the critical fracture energy. 
Here, notice that Eq.~\eqref{eq:stokes} and Eq.~\eqref{eq:darcy} describes the solid and fluid motions in $\mathcal{B}_S$ and $\mathcal{B}_D$, respectively. 
By adopting the effective stress principle, we incorporate the damage in the solid matrix from the phase field evolution equation [Eq.~\eqref{eq:phase_field}] as follows:
%Here, notice that Eqs.~\eqref{eq:eq7} and \eqref{eq:eq8} describe the solid motion, while Eqs.~\eqref{eq:eq9}-\eqref{eq:eq11} governs the fluid flow. 
%By adopting the effective stress principle, we incorporate the damage in the solid matrix from the phase field evolution equation [Eq.~\eqref{eq:eq12}] as follows:
\begin{equation}
\label{eq:effective_stress}
\tensor{\sigma}' = g(d) \tensor{\sigma}_0'^{+} + \tensor{\sigma}_0'^{-},  
\end{equation}
where $g(d) = (1-d)^2$ is the degradation function, and $\tensor{\sigma}_0'^{\pm}$ are the tensile and compressive parts of the fictitious undamaged effective stress \citep{miehe2010phase, mauthe2017hydraulic}. 
In order to enforce crack irreversibility, we introduce a history function $\mathcal{H}$ in Eq.~\eqref{eq:phase_field}, which is a pseudo-temporal maximum of the tensile part of the effective strain energy density ($\psi_e^{+}$) stored in solid skeleton, i.e.,
\begin{equation}
\label{eq:history_variable}
\mathcal{H} = \max_{\tau \in [0, t]} \psi_e^{+}.
\end{equation}
In addition, the undamaged solid matrix in $\mathcal{B}_D$ is considered to be linear elastic, i.e.,
\begin{equation}
\label{eq:linear_elastic}
\tensor{\sigma}_0' = \mathbb{C}^e : \tensor{\varepsilon} \: \: ; \: \: \tensor{\varepsilon} = \symgrad{\vec{u}},
\end{equation}
and fluid flux in $\mathcal{B}_D$ is described by Darcy's law:
\begin{equation}
\label{eq:darcy}
\bar{\vec{w}}_{fD} = - \frac{k}{\mu_f} \grad{p}_{fD},
\end{equation}
where $k$ is the effective permeability of the porous matrix that depends on Kozeny-Carman equation.

The Stokes flow and the Darcy flow can properly be coupled along the interface $\Gamma^*$ by enforcing three transmissibility constraint therein. 
The transmissibility conditions govern the interaction between free and porous-medium flows. 
The first interface condition is the continuity equation that ensures the fluid mass conservation across $\Gamma^*$:
\begin{equation}
\label{eq:interface_eq1}
(\bar{\vec{w}}_{fS} - \bar{\vec{w}}_{fD}) \cdot \vec{n}^* = 0 \: \: \: \text{on } \Gamma^*.
\end{equation}
Here, notice that $\bar{\vec{w}}_{f \alpha} = \phi ( \vec{w}_{f \alpha} - \dot{\vec{u}} ), \alpha = \left\lbrace S, D \right\rbrace$; where $\vec{w}_{f \alpha}$ is the absolute fluid velocity in region $\alpha$ and $\phi$ is the porosity.  
The second condition is the force equilibrium equation that balances the normal stresses across the interface:
\begin{equation}
\label{eq:interface_eq2}
\vec{t}_{fS}^* \cdot \vec{n}^* + p_{fD} = 0 \: \: \: \text{on } \Gamma^*,
\end{equation}
where $\vec{t}_{fS}^*$ indicates the fluid traction vector acting on $\Gamma^*$. 
The third condition is the well-known Beavers-Joseph-Saffman law \citep{beavers1967boundary, saffman1971boundary, mikelic2000interface, layton2002coupling} which correlates the slip velocity with the shear stress along the interface $\Gamma^*$:
\begin{equation}
\label{eq:interface_eq3}
\vec{t}_{fS}^* \cdot \vec{m}_i^* + \mu_f \frac{\alpha_{SD}}{\sqrt{k}}(\bar{\vec{w}}_{fS} - \bar{\vec{w}}_{fD}) \cdot \vec{m}_i^* = 0, 
\: i = \left\lbrace 1,2 \right\rbrace 
\: \: \: \text{on } \Gamma^*,
\end{equation}
where $\alpha_{SD}$ is the dimensionless Beavers–Joseph slippage coefficient which can be determined experimentally \citep{beavers1967boundary}.

Starting from the strong form, we first follow the standard procedure to recover the variational form of Eqs.~\eqref{eq:stokes}-\eqref{eq:phase_field} that contains transmissibility conditions [Eqs.~\eqref{eq:interface_eq1}-\eqref{eq:interface_eq3}] while employing the low-order finite elements for $\vec{u}$, $\bar{\vec{w}}_{fS}$, $p_{fS}$, $p_{fD}$ and $d$. 
By using Eqs.~\eqref{eq:vol_int} and \eqref{eq:surf_int}, we then reformulate the sharp interface variational form into the variational form of the hydromechanically coupled diffuse Stokes-Darcy problem. 
The solution procedure is based on the operator-split scheme to successively update the field variables. 
In this operator-split setting, the phase field $d$ is updated first while all other field variables are held fixed, and then the solver holds the phase field and advances the variables $\vec{u}$, $\bar{\vec{w}}_{fS}$, $p_{fS}$ and $p_{fD}$. 
For brevity, we omit the detailed variational formulation. 
Interested readers are referred to \citet{stoter2017diffuse}, where they adopt similar approach to replace integrals over sharply defined volumes and surfaces by diffuse integrals formulated in terms of phase field.

\section{Numerical example}
\label{numerical_example}

\begin{figure}[h]
\centering
\includegraphics[height=0.325\textwidth]{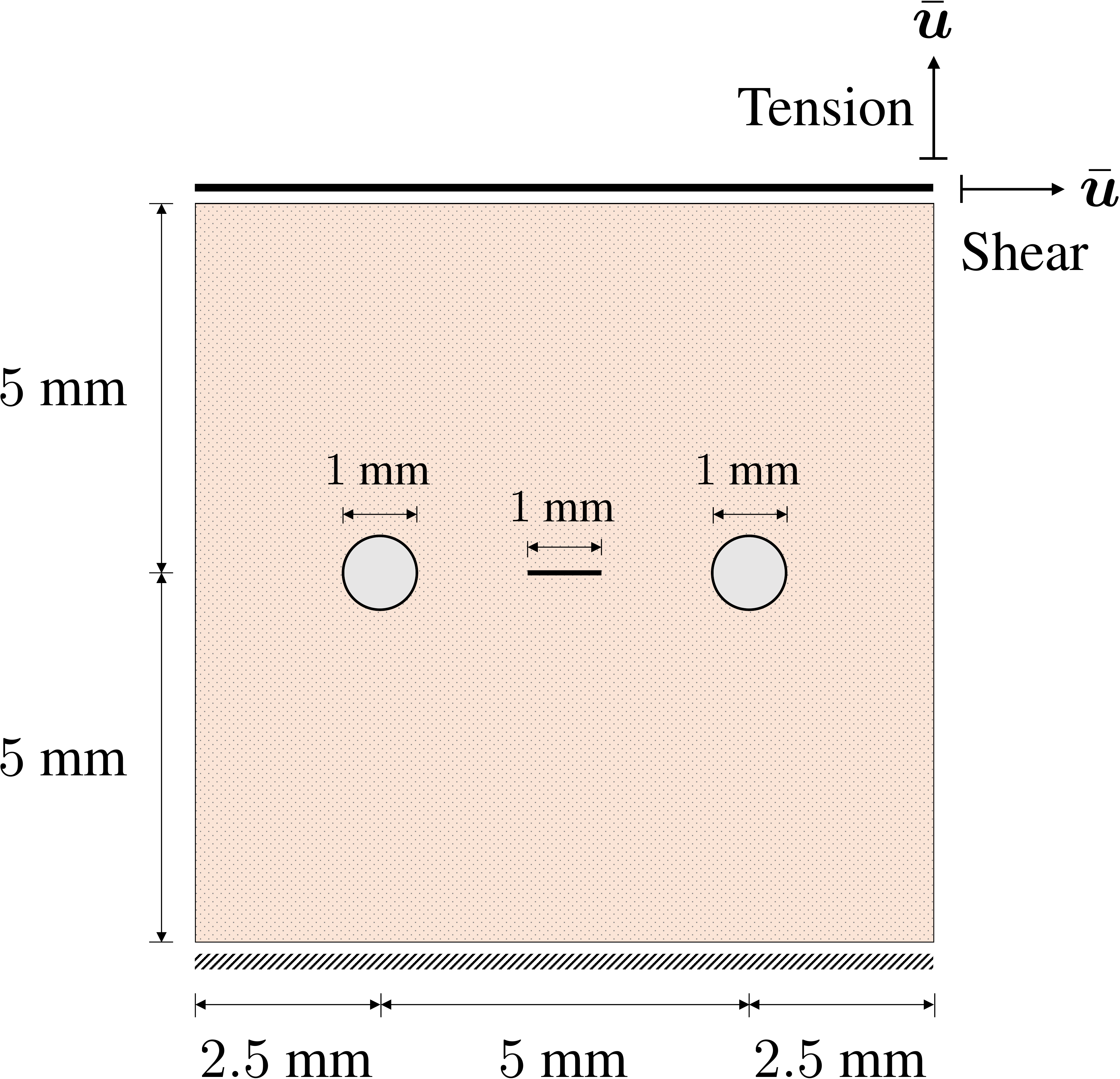}
\caption{Setup of boundary value problems.}
\label{fig:schematic}
\end{figure}

This section presents a numerical example to showcase the applicability of the proposed phase field model for fluid-infiltrating vuggy porous media. 
For simplicity, we limit our attention to two-dimensional case. 
As illustrated in Fig. \ref{fig:schematic}, our problem domain consists of a square-shaped porous matrix that possesses an initial notch and two cavities at the middle. 
The material parameters for this example are chosen as follows: Young's modulus $E = 20$ GPa, Poisson's ratio $\nu = 0.2$, initial permeability $k_{0} = 1.0 \times 10^{-12}$ m$^2$, fluid viscosity $\mu_f = 0.02$ Pa$\cdot$sec, initial porosity $\phi_0 = 0.25$, Beavers-Joseph slippage coefficient $\alpha_{SD} = 0.1$, critical energy release rate $\mathcal{G}_c = 50$ J/m$^2$, and phase field regularization length $l^* = 0.075$ mm. 
Within the same computational domain, two different types of simulations are conducted: the pure tensile test with prescribed vertical displacement $\Delta \bar{u}_y = 0.01$ mm/s; and the pure shear test with prescribed horizontal displacement $\Delta \bar{u}_x = 0.01$ mm/s. 
In both cases, the displacements are prescribed along the entire top boundary, while the bottom part of the domain is fixed. 
Furthermore, zero pore pressure conditions ($\bar{p}_{fD} = 0$) are applied at the top and the bottom, while no-flux boundary conditions are applied at the left and the right boundaries.

\begin{figure}[h]
\centering
\includegraphics[width=0.343\textwidth]{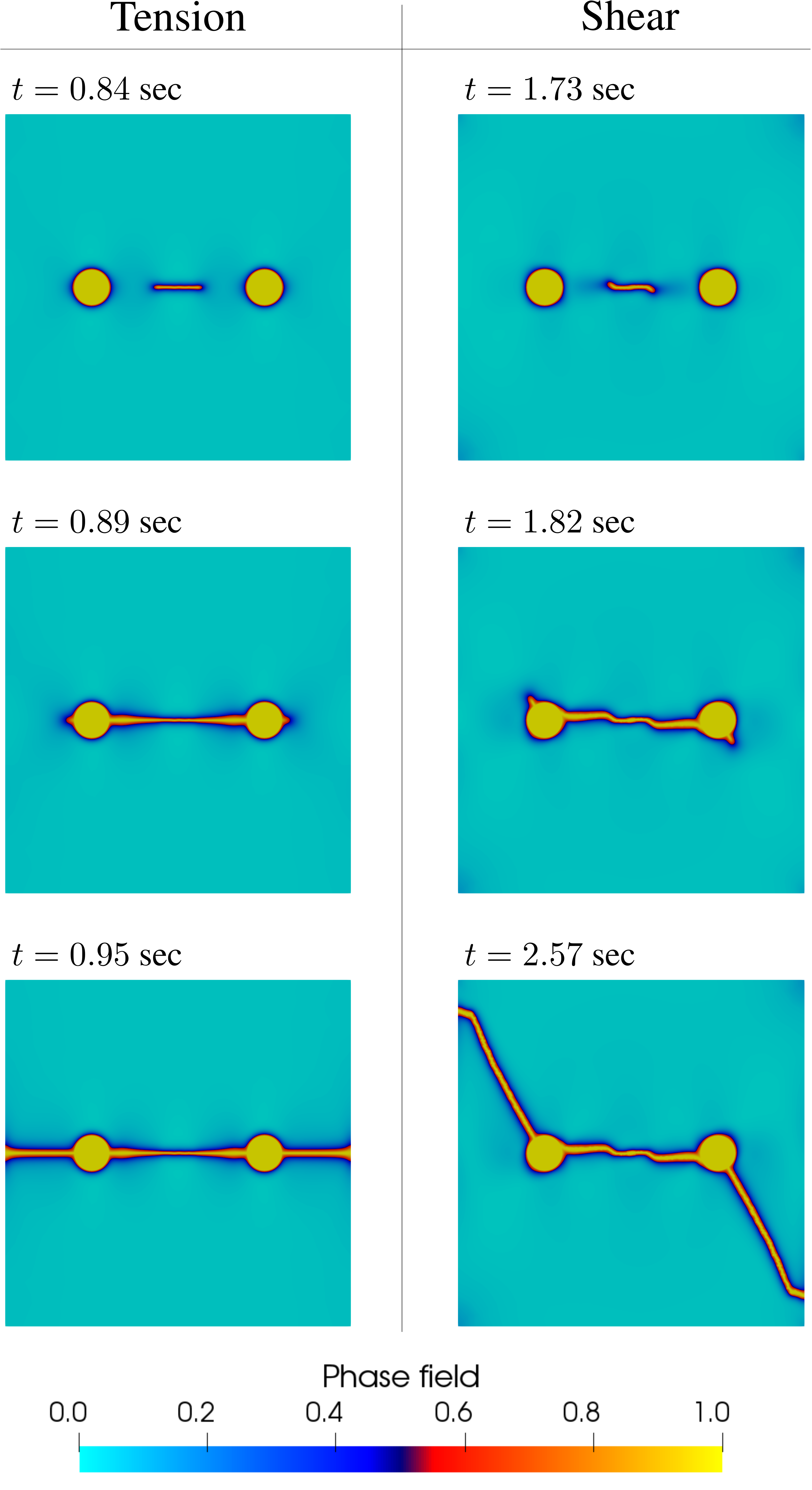}
\caption{Fracture patterns for tension and shear tests.}
\label{fig:frac_phase_field}
\end{figure}

\begin{figure}[h]
\centering
\includegraphics[height=0.4\textwidth]{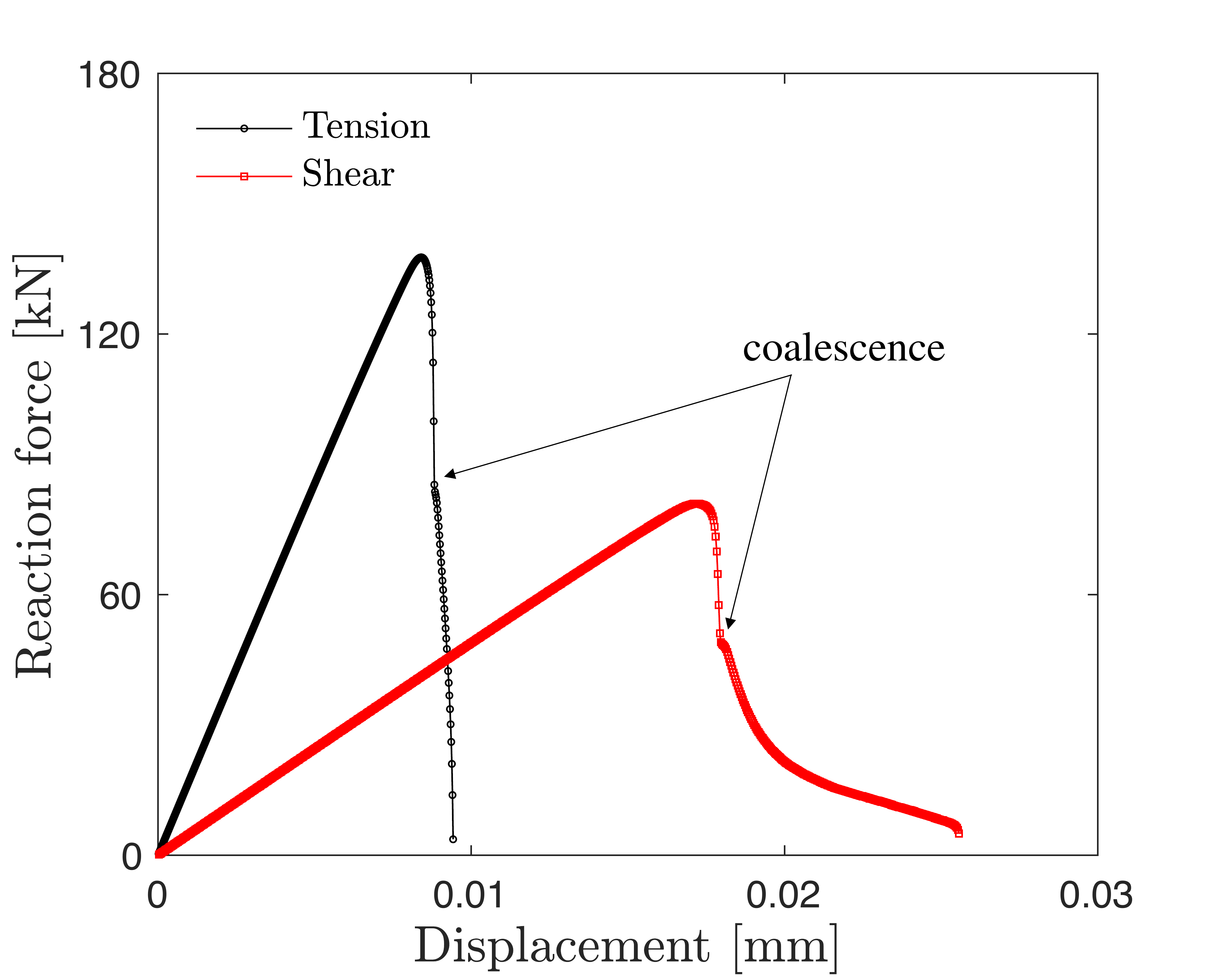}
\caption{The force-displacement curves obtained from tension and shear tests.}
\label{fig:force_disp}
\end{figure}

Fig. \ref{fig:frac_phase_field} illustrates the predicted crack trajectories for both tension and shear tests, while Fig. \ref{fig:force_disp} shows their global force-displacement responses. 
Since notch tip experiences higher stress concentration compared to matrix-cavity interface, cracks tends to initiate from the tips of the initial notch, and then coalescence with cavities. 
Notice that, in pure shear test, inclined crack propagation occurs at the beginning, but eventually coalesces with the cavities due to an increase of the singularity \citep{qinami2019circumventing}. 
After coalescence, pure tensile loading exhibits the horizontal crack pattern due to the symmetry of the boundary value problem, whereas pure shear test yields inclined crack pattern. 

\begin{figure}[h]
\centering
\includegraphics[width=0.343\textwidth]{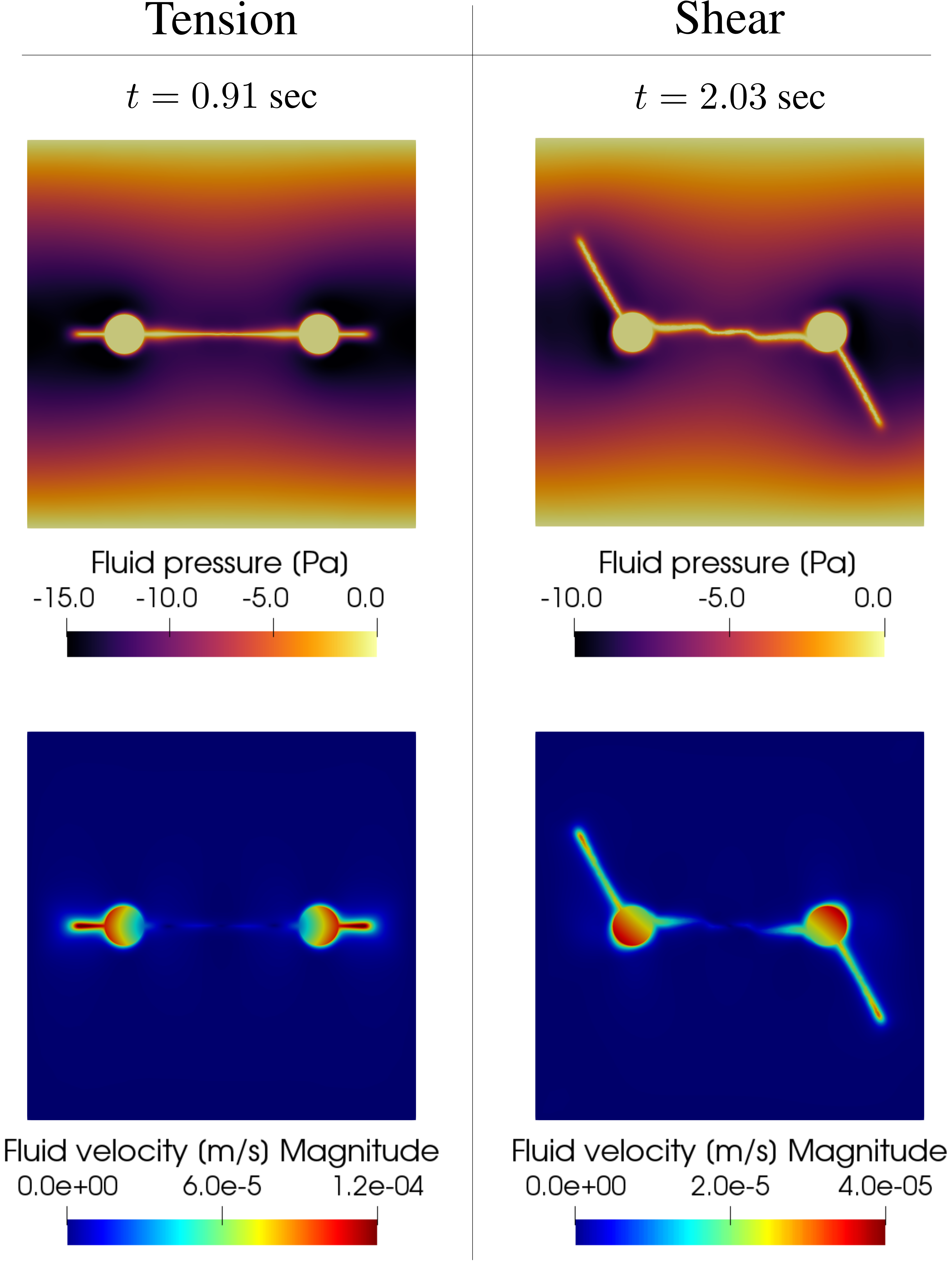}
\caption{Resultant fluid pressure [$p_f = d p_{fS} + (1-d) p_{fD}$] and velocity [$\bar{\vec{w}}_f = d \bar{\vec{w}}_{fS} + (1-d) \bar{\vec{w}}_{fD}$] fields for tension and shear tests.}
\label{fig:pressure_velocity}
\end{figure}

Fig. \ref{fig:pressure_velocity} illustrates the resultant pressure and velocity fields for tension and shear tests at $t = 0.91$ sec and $t = 2.03$ sec, respectively. 
Compared to the permeability enhancement models \citep{mauthe2017hydraulic, heider2017phase}, in our new approach, fluid pressure inside the fracture and cavity remains low while high negative pressure builds up in surrounding Darcy region due to the transmissibility conditions along the interface. 
In addition, compared to the undamaged porous matrix, fluid velocity inside the fracture and cavity exhibits higher values, since it can be interpreted as a region that possesses extremely high permeability. 
The results highlight that our new approach is capable of modeling fracture-cavity interaction in fluid-infiltrating porous media by explicitly modeling the interface conditions.

\section{Conclusion}
\label{conclusion}
This study presented a phase field method for brittle fracture in fluid-infiltrating vuggy porous media. 
To the best of the authors' knowledge, this is the first ever mathematics model that employs the phase field fracture framework combined with coupled Stokes-Darcy flow. 
By adopting the mixture theory, we started with sharp interface formulation with particular emphasis on the interfacial transmissibility conditions, and then transferred it into a diffusive problem by assuming the existence of a phase field representation of the interface. 
We believe that our approach is a significant step towards hydraulic fracturing simulation capability, since this model allows one to explicitly model friction at the crack surface and to explore fracture-cavity interaction without any numerical treatment.

\section{Acknowledgments}
\label{acknowledgements}
The first author is supported by the Earth Materials and Processes
program from the US Army Research Office under grant contract 
W911NF-18-2-0306. 
The second author is supported by 
by the NSF CAREER grant from Mechanics of Materials and Structures program
at National Science Foundation under grant contract CMMI-1846875, the Dynamic Materials 
and Interactions Program from the Air Force Office of Scientific 
Research under grant contracts FA9550-17-1-0169 and FA9550-19-1-0318.
These supports are gratefully acknowledged. 
The views and conclusions contained in this document are those of the authors, 
and should not be interpreted as representing the official policies, either expressed or implied, 
of the sponsors, including the Army Research Laboratory or the U.S. Government. 
The U.S. Government is authorized to reproduce and distribute reprints for 
Government purposes notwithstanding any copyright notation herein.

\bibliographystyle{plainnat}
\bibliography{main}

\end{document}